\pdfoutput=1 % arXiv recommended

\documentclass[11pt]{article}
\usepackage{amsmath}
\usepackage{graphicx}

\usepackage{setspace} % double spacing -- doesn't work with mbe.CLS
\usepackage[hmargin=1in]{geometry}

\usepackage{lineno}

\usepackage[nomargin,inline,index,draft]{fixme}  % Switch draft to final for final
\fxusetheme{color}
\usepackage{natbib}
\bibpunct{(}{)}{;}{a}{}{,} % 5th arg: punctuation between author and year

\DeclareMathOperator*{\argmax}{arg\,max}

\newcommand{\pd}[2]{\frac{\partial  #1}{\partial #2}}

\newcommand{\one}[1]{\mathbf{1}_{#1}}

\begin{document}

\title{Maximum Likelihood Estimation of Frequencies of Known Haplotypes from Pooled Sequence Data}

\author{{%
Darren Kessner$^1$, Tom Turner$^2$,
John Novembre$^{1,3}$
}\\
%\affiliation{$^\ast$Bioinformatics Interdepartmental Program, University of California, Los Angeles;
%$\dagger$Department of Ecology and Evolutionary Biology, University of California, Los Angeles
%}
}

% begin title page (submission)
\maketitle % remove for mbe
\noindent
$^1$Bioinformatics Interdepartmental Program, University of California, Los Angeles \\
$^2$ Department of Ecology, Evolution, and Marine Biology, University of California, Santa Barbara\\
$^3$Department of Ecology and Evolutionary Biology, University of California, Los Angeles \\

\noindent
Submission:  Research Article \\
Institution where work was done: UCLA \\
Keywords: maximum likelihood, EM algorithm, haplotype frequency estimation, pooled sequence data \\
Running Head: Haplotype Frequency Estimation from Pooled Sequence Data \\

\noindent
Corresponding author: \\
John Novembre \\
Terasaki Life Sciences Building \\
610 Charles E. Young Drive, East \\
Los Angeles, CA 90095-7239 \\
jnovembre@ucla.edu \\
% end title page

\newpage
%\linenumbers

%\authorlist{Kessner and Novembre}
%\shorttitle{Haplotype Frequency Estimation from Pooled Sequence Data}

%\mbeabstract{% 
\section*{Abstract} % remove for mbe
DNA samples are often pooled, either by experimental design, or because the
sample itself is a mixture. For example, when population allele frequencies are
of primary interest, individual samples may be pooled together to lower the
cost of sequencing. Alternatively, the sample itself may be a mixture of
multiple species or strains (e.g. bacterial species comprising a microbiome, or
pathogen strains in a blood sample). We present an expectation-maximization
(EM) algorithm for estimating haplotype frequencies in a pooled sample directly
from mapped sequence reads, in the case where the possible haplotypes are
known.  This method is relevant to the analysis of pooled sequencing data from
selection experiments, as well as the calculation of proportions of different
strains within a metagenomics sample.  Our method outperforms existing methods based
on single-site allele frequencies, as well as simple approaches using sequence
read data.  We have implemented the method in a freely available open-source
software tool.
%}

%\keywords{maximum likelihood, EM algorithm, haplotype frequency estimation, pooled sequence data}

%\email{dkessner@ucla.edu}

%\mbestyle

\section*{Introduction}

Pooled sequencing is a common experimental method in which DNA samples from
multiple individuals are sequenced together.  In some contexts, the pooling of
individual samples is performed by the researcher; in others, the sample itself
is a mixture of multiple individuals.  When population allele frequencies are
of primary interest, pooled sequencing approaches can reduce the cost and labor
involved in sample preparation, library construction, and sequencing
\citep{Futschik2010, Cutler2010, Kofler2011, Orozco-terWengel2012, Huang2012}.

For example, in experimental evolution studies, populations are selected for
extreme values of a trait over several generations, followed by pooled
sequencing to calculate allele frequencies at polymorphic sites across the
genome \citep{Nuzhdin2007, Burke2010, Earley2011, Turner2011, Zhou2011}.
Typically, differences in single-site allele frequencies between an experimental population and a
control population (or between two experimental populations selected in
opposite directions) are used to identify regions of the genome that may have
undergone selection during the course of the experiment, and thus contribute to
the trait of interest.  However, localizing such regions would be improved if
haplotype frequencies were more easily able to be estimated from pooled data, 
as many of the most powerful tests for selection rely on haplotype information 
\citep{Voight2006, Sabeti2007}. 

In certain cases, haplotype frequency estimation may be more feasible than others, such as when the investigator has prior knowledge about the founders of the pooled sample.  For example, \citet{Turner2012} used inbred
lines from the Drosophila Genetic Reference Panel (DGRP) \citep{Mackay2012} to
create the founding population for the selection experiment.  In such an
experiment, individual haplotypes in the evolved populations will be, apart
from \emph{de novo} mutations, mosaics of haplotypes from the founding
population, whose sequences are known.  This structure should make it simpler to estimate haplotype frequencies, and in turn detect regions harboring adaptive variation, by searching for haplotypes that have increased in frequency locally during the experiment. 
 
In many other contexts, biological samples are naturally pooled, and the
researcher is interested in the relative proportions of various species or
strains within the sample.  For example, malaria researchers interested in drug
resistance and vaccine efficacy testing have developed several laboratory and
computational techniques for determining the proportions of different malaria
parasite strains in blood samples \citep{Cheesman2003, Hunt2005, Takala2006,
Li2007, Hastings2008, Hastings2010}.  In metagenomics studies, it is common to
compare the microbiota proportions of multiple individuals, or of different
tissues or locations within a single individual \citep{Ley2006}.  In these examples,
the canonical haplotypes of the strains are known, and it is the relative
frequencies that are of interest.

Indirect estimation of haplotype frequencies from unphased genotype data
has a long history (see \cite{Niu2004} for a review of these methods).
Several approaches for estimating haplotype frequencies from pools containing multiple
individuals have focused on the use of single-nucleotide polymorphism
(SNP) allele frequencies obtained by array-based genotyping 
\citep{Pe'er2003, Ito2003, Wang2003, Yang2003, Kirkpatrick2007, Zhang2008, Kuk2009}.  
Some examples of this class of methods have incorporated prior knowledge
about haplotypes in the sample into the estimation \citep{Gasbarra2009, Pirinen2009}.  
Most recently, \citet{Long2011} have proposed a method for estimating haplotype
frequencies from SNP allele-frequency data obtained by pooled sequencing, using
a regression-based approach with known haplotypes.

Pooled sequence data provide two important sources of information beyond single-site allele
frequencies:  haplotype information from sequence reads that span multiple
variant sites, and base quality scores, which give error probability estimates
for each base call.  Here we introduce a method to use this additional
information to estimate haplotype frequencies from pooled sequence data, in the
case where the constituent haplotypes are known.  This method uses a
probability model that naturally incorporates uncertainty in the reads by using
the base quality scores reported with the sequence data.  The method obtains a
maximum likelihood estimate of the haplotype frequencies in the sample via an
expectation-maximization (EM) algorithm \citep{Dempster1977}.  We present
results from realistic simulated data to show that the method outperforms
allele-frequency based methods, as well as simple approaches that use sequence
reads.  The use of a fixed list of known haplotypes allows the algorithm to use
data from much larger genomic regions than algorithms that enumerate all
possible haplotypes in a region, which leads to much improved haplotype
frequency estimates.  We have implemented the method in an open-source software
tool \texttt{harp} (see authors' websites for software link).

\section*{Methods}

\begin{figure}[h]
\begin{center}
\includegraphics[scale=0.5]{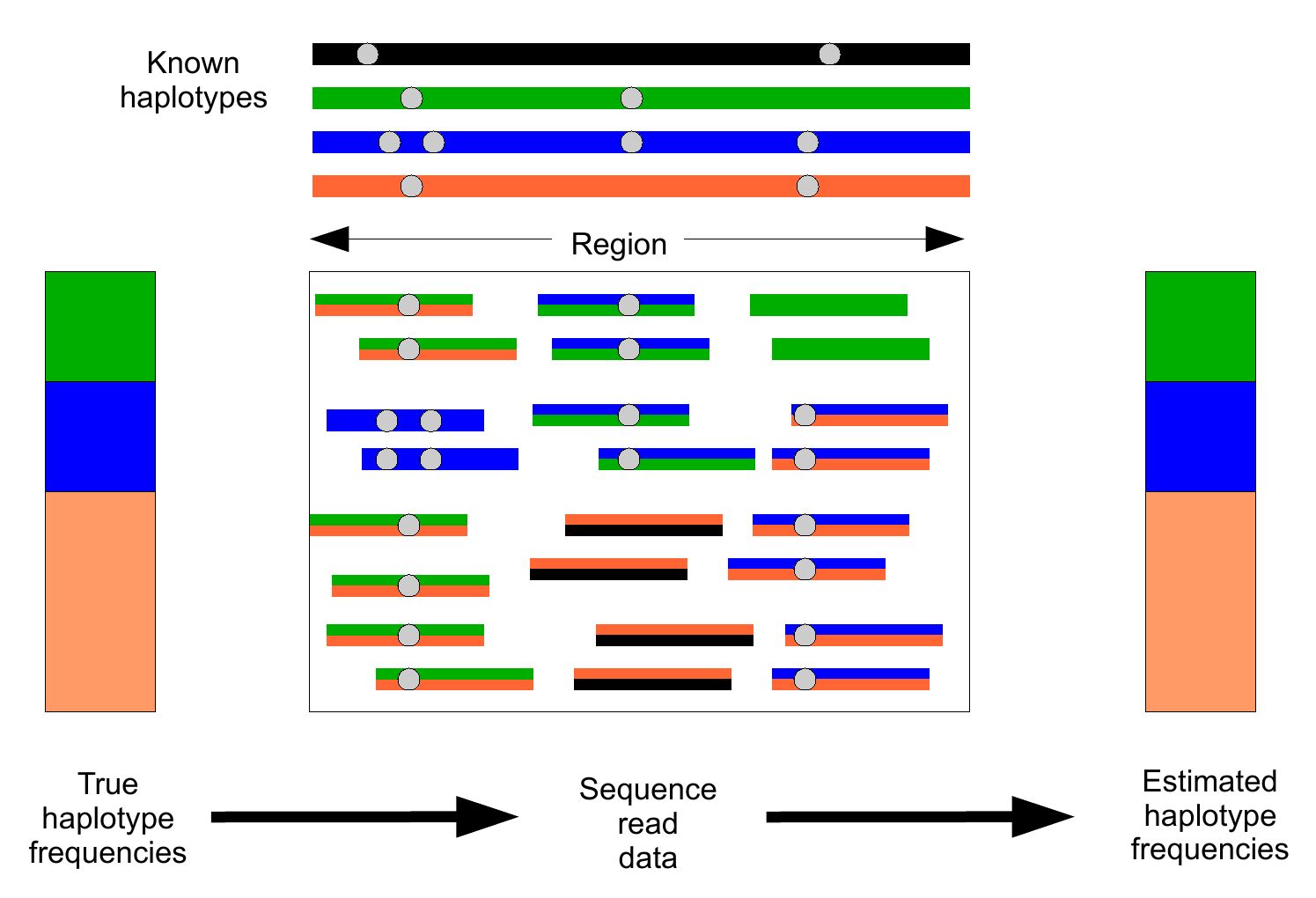}
\end{center}
\caption{Haplotype information from individual reads can be combined across a
    genomic region to obtain haplotype frequency estimates.  In this cartoon,
    there are 4 known haplotypes (black, green, blue, orange), with sequence
    data coming from a pool containing 25\% green, 25\% blue, and 50\% orange
    haplotypes.  Each read is probabilistically assigned to the known
    haplotypes.  Some reads can be assigned with great certainty, e.g
    the reads coming from the blue haplotype that cover two neighboring variant
    sites.  Other reads (represented by two colors) are assigned with less
    certainty.}
\label{em_single_window}
\end{figure}

We assume that there are $H$ haplotypes represented in the pool, and that the
sequence reads have been generated randomly according to the frequencies of the
haplotypes.  Informally, we use haplotype information contained in an
individual read to probabilistically assign that read to one or more of the
known haplotypes (Figure \ref{em_single_window}), and then use the
probabilistic haplotype assignments to estimate the haplotype frequencies.

\subsection*{Probability Model}

Let $f = \left(f_1, \ldots, f_H \right)$ 
be the frequencies of the $H$ haplotypes in the genomic region of interest.
We can think of $N$ sequence reads $r = \left(r_1, \ldots, r_N \right)$ 
as being independently generated as follows.
To generate read $r_j$: 
\begin{itemize}    
    \item choose the haplotype $\eta_j$ to copy from: \\ $\eta_j \sim \text{Discrete(f)}$
    \item choose a starting position uniformly at random in the genomic region, and
          copy read $r_j$ from haplotype $\eta_j$ starting at the chosen position
    \item draw base quality scores for the read from a fixed distribution (which can be determined
          empirically)
    \item introduce errors in the sequence read, with the probability of error in a base call
          given by the base quality score at that position
\end{itemize}

In practice, haplotypes may not be perfectly known, or there may be segregating
variation within the strain represented by a particular haplotype.  In such
cases, International Union of Pure and Applied Chemistry (IUPAC) ambiguous base
codes (e.g. R for purine, Y for pyrimidine, N for any, etc.) may be used in
place of the standard bases (A, C, G, T) to indicate the uncertainty.  We
incorporate these cases into our probability model by assuming the true base at each segregating site is sampled from a discrete distribution with probabilities determined by the allele frequencies at that site within the strain (which may be known a priori or assumed to be uniform).  

\subsection*{Haplotype Likelihood}

Calculating the likelihood of a set of haplotype frequencies given read data under this model can be carried out as follows.  Let $L_j$ be the length of the $j^{\, th}$ read $r_j$, let $(r_j[1], \dots,
r_j[L_j])$ be the base calls, and let $q_j = (q_j[1], \dots, q_j[L_j])$ be the
base quality scores.  Also, let $(\eta_j[1], \dots, \eta_j[L_j])$ be the
corresponding bases of haplotype $\eta_j$.  At read position $i$, $q_j[i]$ is
the probability of sequencing error at that position:  
$q_j[i] = P(r_j[i] \neq \eta_j[i])$.  Note that for paired-end data, $r_j$
represents a read pair coming from a single haplotype, and that the positions
within the read may not be contiguous.

We have $P(\eta_j, r_j | f, q_j) = P(\eta_j | f) P(r_j | \eta_j, q_j)$.   The
first term $P(\eta_j | f)$ is given by the discrete distribution 
with probabilities $f$, and the second term $P(r_j | \eta_j, q_j)$, the \emph{haplotype
likelihood}, can be calculated from the base quality scores, as follows. 

First, we assume that sequencing errors within a single read are independent of each other:
\begin{align*}
    P( r_j \, | \, \eta_j, q_j ) &= \prod_{i=1}^{L_j} P(\,r_j[i] \;|\; \eta_j[i], q_j[i] )
\end{align*}

Next, we need to specify how to calculate the terms in the above product, i.e.
the probability of an observed base, given the true base and the base quality
at that position.  For simplicity, we assume that each of the 3 incorrect bases
will be observed with equal probability:
\begin{align*}
    P(\,r_j[i] \; | \; \eta_j[i] , q_j[i]) &= \left\{ \begin{matrix} 
                                1 - q_j[i] \quad  &\text{if  } r_j[i] = \eta_j[i]  \\ 
                                q_j[i] / 3 \quad  &\text{if  } r_j[i] \neq \eta_j[i]
                            \end{matrix} \right.
\end{align*}
More generally, we note that we can use a base error matrix
(parametrized by base quality score) to allow for unequal probabilities, and
that these probabilities can be estimated from the data by considering the
monomorphic sites in the sample.  

Note that if position $i$ is a segregating site in the strain represented by haplotype $\eta_j$,
the likelihood is calculated by summing over the possible bases:
\begin{eqnarray*}
    \lefteqn{P(\,r_j[i] \; | \; \eta_j[i], q_j[i]) =} \\
& \sum_{b \in \{A,C,G,T\}} P(\,r_j[i] \; | \; \eta_j[i] = b, \, q_j[i]) \; P(\eta_j[i] = b)
\end{eqnarray*}
where $P(\eta_j[i] = b)$ is the frequency of base $b$ at that site within the strain.  For sites
where the possible bases are known, but not the allele frequencies, we set the allele frequencies
to be equal, e.g. .5 for biallelic sites, and .25 for sites with no information.

For clarity, we suppress the dependence on the base quality scores in what follows.

\subsection*{Simple Approaches}

We explored two simple approaches for estimating haplotype frequencies.  The
first method is a simple string match algorithm, where sequence reads are
fractionally assigned (with equal weight) to haplotypes with which they
are identical up to a specified maximum number of mismatches.  For example, a
read that matches two haplotypes is assigned .5 to each.  The fractional 
assignments are then summed, to obtain counts for each haplotype, and dividing
by the number of reads gives the haplotype frequency estimate.

The second method, which we call a \emph{soft} string match, uses the
probability model described above to calculate the vector of haplotype
likelihoods $l_j$ for each read $r_j$.  Thus, the soft string match makes use
of the base quality scores from the reads.  The haplotype likelihood vector
$l_j$ is normalized so that the components sum to 1, which we take to be our
probabilistic haplotype assignment.  As with the fractional assignments above, the
probabilistic assignments are averaged to obtain the haplotype frequency
estimate.

\subsection*{EM Algorithm}

In addition to the simple approaches, we developed a full likelihood approach to obtain
maximum likelihood estimates of the haplotype frequencies under the 
probability model described above.

We assume that our reads are generated independently, so our complete data likelihood is:
\begin{equation*}
    L(f \, | \, \eta,r) \:=\: P(\eta,r \, | \, f) \:=\: \prod_{j=1}^N P(\eta_j, r_j | f)
\end{equation*}

We observe the reads $r$, but treat the haplotype assignments $\eta$ as missing
data, so we are interested in the marginal likelihood,
\begin{equation*}
L(f \, | \, r) \:=\: P(r \, | \, f) \:=\: \sum_\eta P(\eta,r|f)
\end{equation*}
which we maximize by iteratively calculating haplotype frequency estimates via the 
EM algorithm: $f^{(0)}, f^{(1)}, \ldots$

First we describe the iteration step of the algorithm; we assume we have $f^{(i)}$ and show how to obtain $f^{(i+1)}$.  
In the appendix, we show that this is the formal EM algorithm of \citet{Dempster1977}.

We let $l_{j,h} = P(r_j|\eta_j=h)$, and let 
$l_j = \left( l_{j,1}, \ldots, l_{j,H} \right)$ 
be the vector of haplotype likelihoods for read $j$.  Note that the haplotype
likelihood vectors can be calculated once and cached, before the actual EM
iteration.

Given $f^{(i)}$, we define $p_j = \left( p_{j,1}, \ldots, p_{j,H} \right)$ 
to be the haplotype posterior vector for read $j$, where 
\begin{equation*}
p_{j,h} \:=\: P(\eta_j = h|r_j, f^{(i)})
\end{equation*}
Intuitively, $p_j$ is a probabilistic haplotype assignment of read $r_j$, with each component $p_{j,h}$
representing the probability that the read came from haplotype $h$ 
(given our current haplotype frequency estimate $f^{(i)}$).
Note that:
\begin{align*}
    P(\eta_j = h|r_j, f^{(i)}) & \propto P(r_j| \eta_j = h) \: P(\eta_j = h|f^{(i)}) \\
                               & = l_{j,h} \: f^{(i)}_h
\end{align*}
so $p_j$ can be obtained by taking the component-wise product $l_j \circ f^{(i)}$, and normalizing so that
the vector components sum to 1.  As a special case, if $f^{(0)}$ is uniform, then in the first iteration, 
$p_j$ is just $l_j$ normalized.

Our updated estimate $f^{(i+1)}$ is given by the average of the haplotype posterior vectors:
\begin{equation*}
    f^{(i+1)} = \frac{\sum_j p_j}{N}
\end{equation*}

Finally, we must specify how to choose our initial haplotype frequency estimate
$f^{(0)}$, as well as convergence criteria for the iteration.  For our first
initial estimate we use the uniform distribution $f^{(0)}_h = 1/H$.  We also
use additional random initial estimates drawn from a symmetric Dirichlet
distribution to start multiple runs of the algorithm, since there is a possibility that
the EM algorithm will climb to a non-global local maximum on the likelihood surface.  For
the termination condition, we specify a threshold $\epsilon$, and halt the
iteration when the squared distance between estimates falls below the
threshold: $ | f^{(i+1)} - f^{(i)} | ^2 < \epsilon $.  
In practice, we found a value of $\epsilon = 10^{-8}$ to work well and this 
value is used in the results presented below.

\subsection*{Base Quality Score Recalibration}

We observed inconsistencies between the reported base quality scores in our
experimental data sets and empirical error rates based on sequence reads
covering monomorphic sites in the known haplotypes (see Results), which motivated the development of a
recalibration method to correct for these inconsistencies.

Illumina base quality scores have different interpretations, depending on the
Illumina version.  In our experimental data set, corresponding to Illumina
versions 1.5 - 1.7, the scores range from 2 to 40, with the score $q$
representing an error probability given by the Phred scale:
\begin{align*}
    P(\text{error}) = 10^{-q/10}
\end{align*}
For example, a base quality score of 20 gives an error probability of 1/100.  
The special score of 2 indicates that the base should not be used in downstream analysis.

To recalibrate, we examine monomorphic sites to calculate an observed error rate
$P_{obs}(\text{error})(q)$ for each possible base quality score $q$.  
These observed error rates can then be used directly in the haplotype
likelihood calculation in place of the Phred scale error rates, or to create a
new BAM file with recalibrated base quality scores.

\subsection*{Implementation}

We implemented both simple approaches and the EM algorithm described above in a
C++ program called \texttt{harp} (\emph{Haplotype Analysis of Reads in Pools}).
The program takes as input a standard BAM file with mapped sequence reads, a
reference sequence in FASTA format, and known haplotypes in the SNP data format
used by the Drosophila Genetic Reference Panel (DGRP) project
\citep{Mackay2012}.  Support for other data formats is currently under
development.  The software uses the \texttt{samtools} API for random access to
BAM files \citep{Li2009}.  The program includes many options for the user to
customize the analysis, including choice of algorithm, the genomic region to
analyze, parameters for sliding windows within the region, convergence
threshold for the EM algorithm, parameters used to generate multiple random
initial estimates to avoid local maxima, and base quality score recalibration.  
The program also calculates standard errors for the haplotype frequency estimates,
using general properties of the EM algorithm and maximum likelihood estimators
(see the Appendix for details on this calculation).

\subsection*{Evaluation}

To evaluate the performance of the algorithms, we used simulated pooled
sequence data based on experimental data from selection experiments in  \emph{Drosophila melanogaster}
\citep{Turner2012}.  The data consisted of Illumina 85bp and 100bp
paired-end sequence reads generated from 4 pools of 120 \emph{D. melanogaster}
individuals each, sequenced at 200x average coverage.  For our known
haplotypes, we used the publicly available SNP data from 162 Drosophila inbred
lines representing Freeze 1 of the DGRP project.  The published haplotypes
include ambiguous base codes (e.g. R for A or G) to represent sites that have
multiple alleles still segregating within the inbred line.  The ambiguous base
code N is used at sites where there was not enough sequence data to make a base
call.

We used the following procedure to simulate pooled sequence data:
\begin{itemize}    
    \item Generate a haplotype frequency distribution (the \emph{true}
        distribution) from a symmetric Dirichlet distribution, with single
        parameter $\alpha$ chosen to produce frequency distributions similar to
        those observed in the real data ($\alpha = .2$).  
    \item Draw random paired-end sequence reads by choosing the haplotype according
        to the true distribution, the starting position uniformly at random
        over the given genomic region, and the paired end distance according to
        a Poisson distribution fitting the real data.
    \item For segregating sites denoted by ambiguous base codes, draw allele
        frequencies according to a symmetric Dirichlet distribution.  Choose
        the true base at a segregating site according to the allele
        frequencies.  (For biallelic sites denoted by a 2-base ambiguous code,
        e.g. R for A or G, we set the Dirichlet parameter $\alpha = 1$, i.e.
        the allele frequency was chosen uniformly at random.  For sites denoted
        by N (any) in the haplotype, we set $\alpha = .1$, as we expect that most of
        these sites have an allele that is at or near fixation).      
    \item Generate base quality scores according to the empirical distribution
        obtained from the real data, and introduce sequencing errors with error
        rates determined by the base quality scores.
\end{itemize}    

Algorithm performance was evaluated by calculating the sum of squared errors
between the true haplotype frequencies used to simulate the data and the
frequencies estimated by the EM algorithm: $\sum_{h=1}^H \left( f^{true}_h -
f^{estimated}_h \right) ^2$.

\section*{Results}

\subsection*{Comparison With Existing Allele-Frequency Based and Simple Sequence Based Methods}

We first evaluated the performance of the EM algorithm in comparison to
single-site allele-frequency based methods and the simple sequence-based methods
discussed above (see \emph{Methods}).  To represent the allele-frequency based
methods, we chose \texttt{hippo}, which is a freely available program that has
been shown to outperform other allele-frequency based methods for estimating
haplotype frequencies \citep{Pirinen2009}.  One property of this class of
methods is that all possible haplotypes in the region are considered during the
estimation.  This results in an exponential growth in the number of haplotypes
(and thus memory usage and algorithm running time) as the region width
increases.  To improve performance, the \texttt{hippo} method allows one to specify known haplotypes, which we do here.  We found it difficult to obtain results on our simulated
data for regions larger than about 2kb (though this distance scale is driven largely by the relatively high Drosophila-specific levels of diversity we simulate here).  

\begin{figure}[h]
\begin{center}
\includegraphics[scale=0.45]{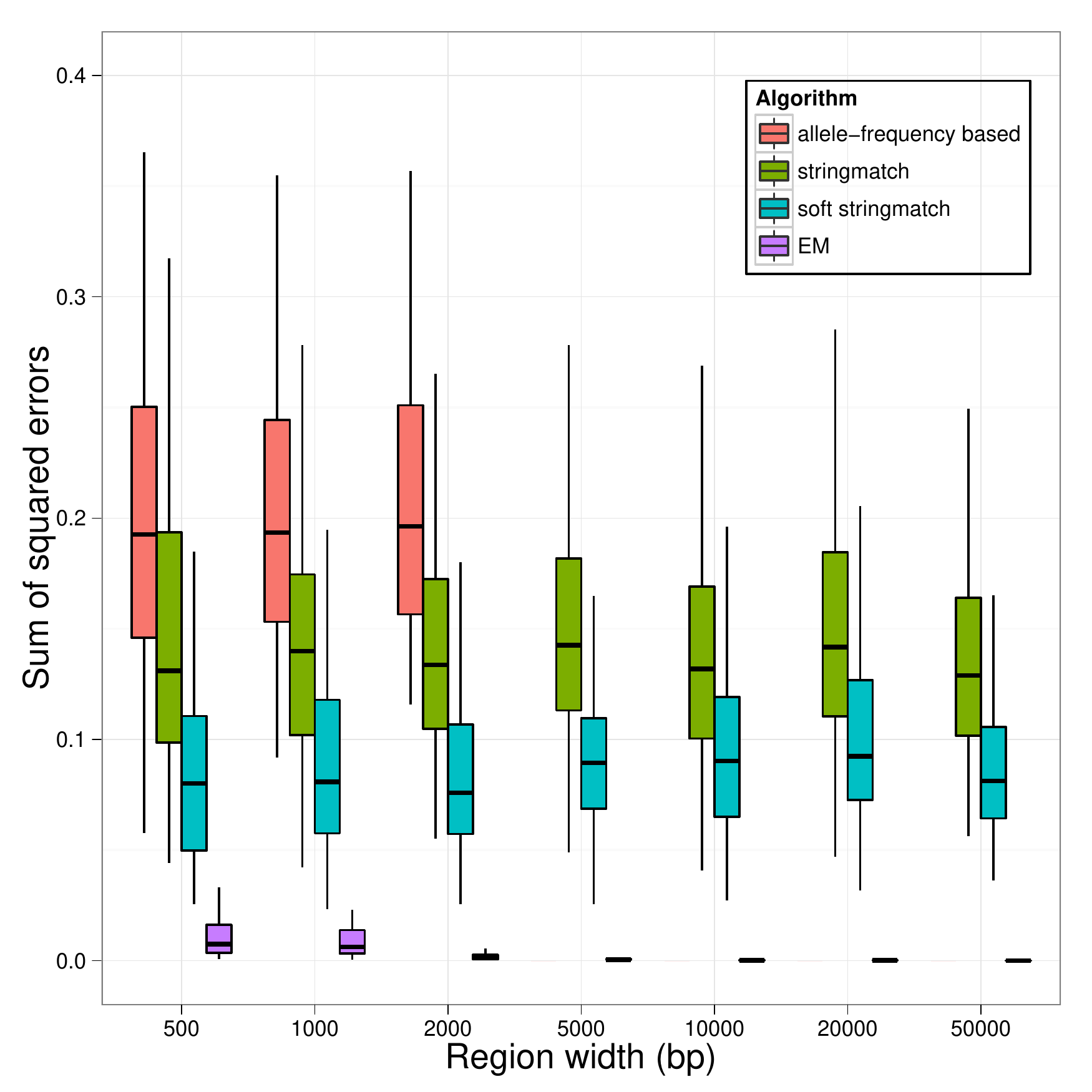}
\end{center}
\caption{Comparison of the EM algorithm to known allele-frequency based and simple sequence-based methods.
         Each algorithm was run on simulated pooled 100bp paired-end sequence data from 20 haplotypes at 200x coverage, with
         100 replicates for each region width.}
\label{algorithm_comparison}
\end{figure}

In this comparison, we simulated data from a pool of 20 haplotypes with 100bp paired-end sequence
reads and 200x pooled coverage, with 100 replicates each from
genomic regions ranging in size from 500bp to 50kb.  

We found that the simple methods using sequence reads outperformed the method
based on single-site allele frequencies, and that the EM algorithm performed
vastly better than all of the other methods (Figure
\ref{algorithm_comparison}).  The soft stringmatch method showed a distinct
improvement over the stringmatch method, due to the incorporation of
information from the base quality scores.

The EM algorithm's increased performance can be attributed to the sharing of
information across all of the reads in the genomic region.  In contrast to the
other methods, the EM algorithm's performance improves as the width of the
region increases.  This improvement comes from the fact that more variant sites
are available to distinguish between haplotypes, in addition to the increased
amount of data on which to base the inference.

\subsection*{Effects of Region Width, Coverage, and Read Length}

\begin{figure}[h]
\begin{center}
\includegraphics[scale=0.45]{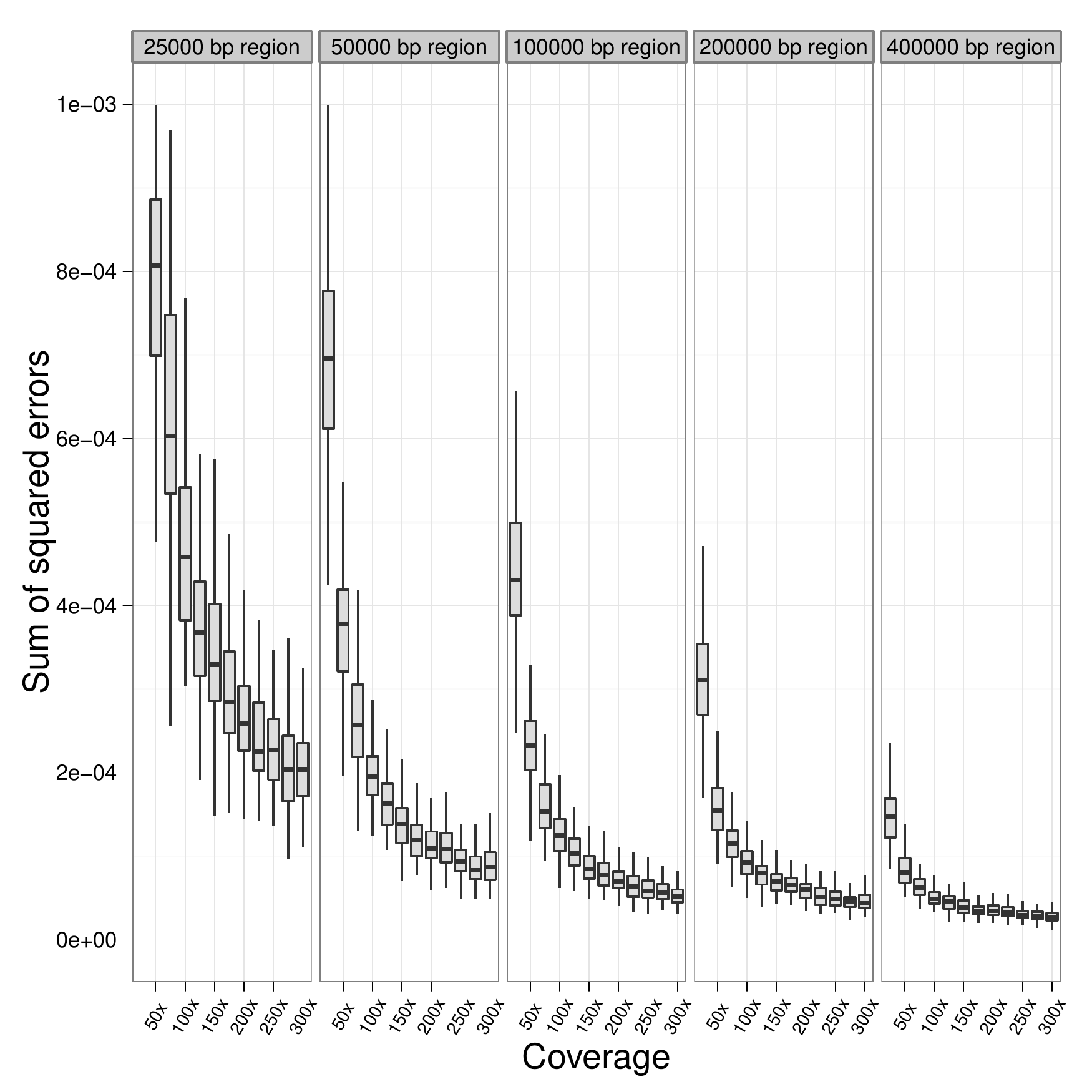}
\end{center}
\caption{Performance of the EM algorithm increases with both coverage and
region width.  The algorithm was run on simulated pooled 100 bp paired-end
sequence data from 162 haplotypes, with 100 replicates for each region width
and coverage level.}
\label{coverage}
\end{figure}

We next evaluated the performance of the EM algorithm with respect to
increasing region width and coverage.  In this evaluation, we simulated pooled
data (100bp paired-end) from all 162 haplotypes, 100 replicates each in genomic
regions ranging in size from 25kb to 400kb, at coverages ranging from 25x to
300x.  We found that performance increases substantially as coverage increases,
especially at the lower coverage levels (25x - 100x), and also as the region
width increases (Figure \ref{coverage}).  In particular, for larger regions ($\geq$
100kb) at moderate pooled coverage (200x), the sum of squared errors is less than
$10^{-4}$, which corresponds to a root mean squared error of less than a tenth
of a percent per haplotype.

\begin{figure}[h!]
\begin{center}
\includegraphics[scale=0.45]{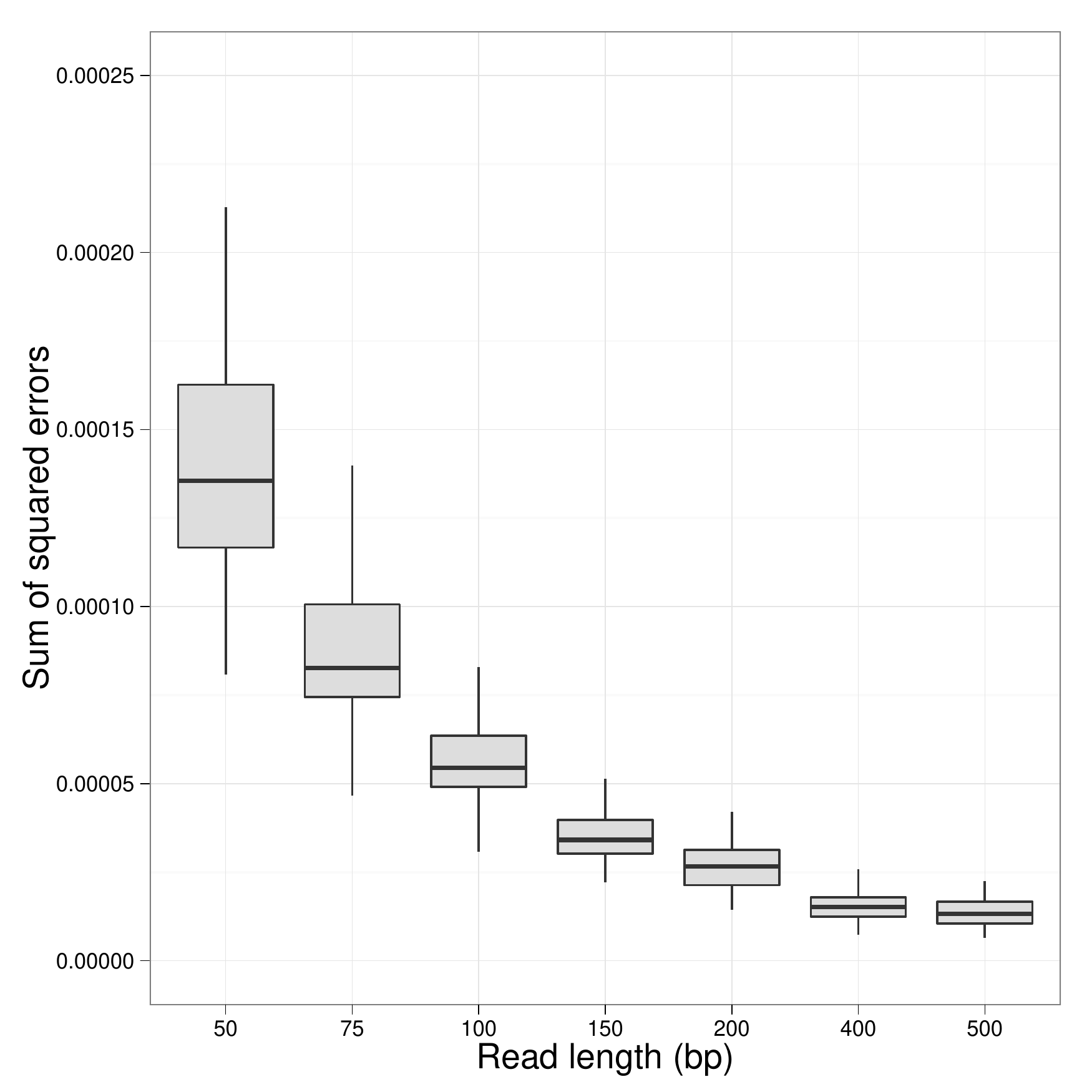}
\end{center}
\caption{The EM algorithm performs better with longer reads, which provide more haplotype information.  
         The algorithm was run on simulated pooled paired-end sequence data from 162 haplotypes
         in a 200kb region, with 100 replicates for each read length.}
\label{read_length}
\end{figure}

We also evaluated the effect of increasing read lengths on the performance of
the EM algorithm.  We simulated paired-end sequence data in a 200kb region with
sequence read lengths ranging from 50bp to 500bp (100 replicates each).  In each case, we generated
200,000 read pairs (200x coverage for 100bp reads).  As expected, longer read
length also increases performance, due to the additional haplotype information
contained in individual reads (Figure \ref{read_length}).

\subsection*{Robustness to Sequencing Errors}

\begin{figure}[h]
\begin{center}
\includegraphics[scale=0.45]{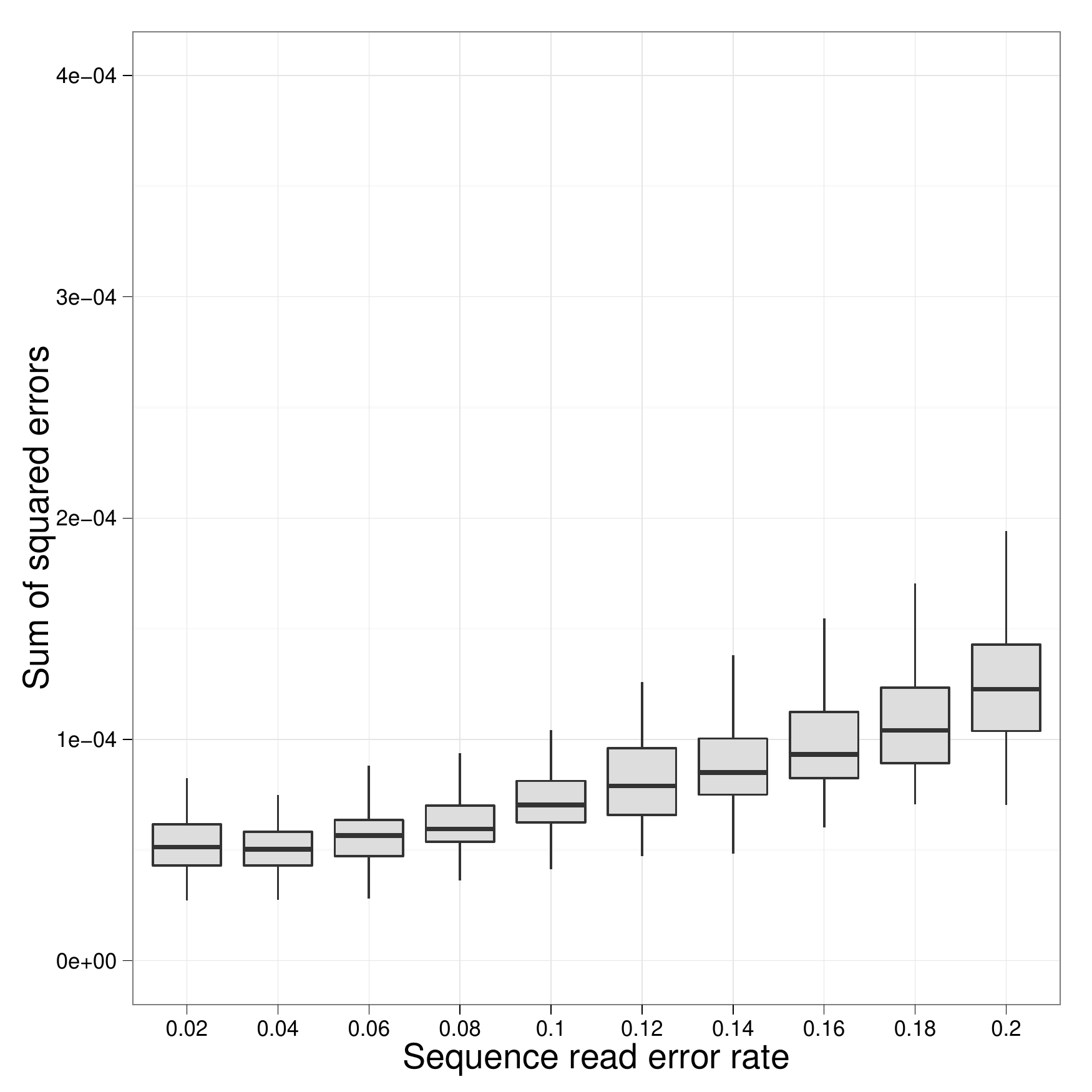}
\end{center}
\caption{The EM algorithm maintains good performance with increasing sequence read error rate.  Empirical
         error rates were found to be in the range of .05 - .07 errors per base call.
         The algorithm was run on simulated pooled 100bp paired-end sequence data from 162 haplotypes
         in a 200kb region, with 100 replicates for each error rate.}
\label{error_rate}
\end{figure}

Next, we studied the effects of sequence read errors on the haplotype frequency
estimation.  We calculated an empirical base quality score distribution, which
we shifted to obtain simulated data sets with specified error rates.  In our
experimental data sets, the sequence error rate calculated from the base
quality scores was generally in the range of $.05 - .07$ (errors per base
call), depending on the region.  On simulated data sets (162 haplotypes, 200kb
region, 100bp paired-end reads, 200x coverage), we found that the EM algorithm
maintains good performance (sum of squared errors $\approx 10^{-4}$), even with
error rates of 2-3x empirical error rates (Figure \ref{error_rate}).

\subsection*{Effects of Haplotype Diversity}

\begin{figure}[h]
\begin{center}
\includegraphics[scale=0.45]{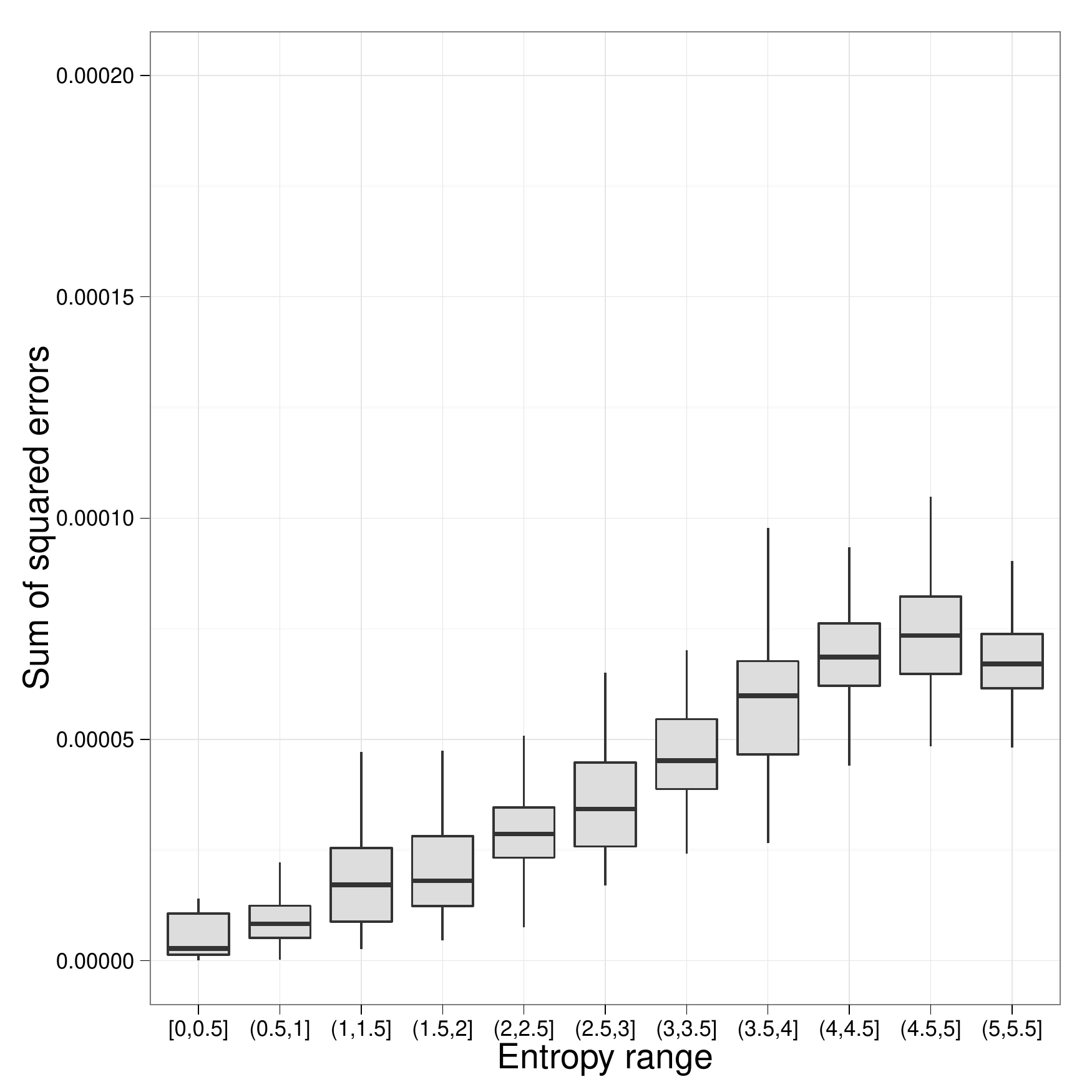}
\end{center}
\caption{The EM algorithm performs best when the true frequency distribution has low entropy.
         The algorithm was run on simulated pooled 100bp paired-end sequence data from 162 haplotypes
         at 200x coverage in a 200kb region (550 replicates binned by Shannon entropy in natural log units).}
\label{entropy}
\end{figure}

We investigated the effects of haplotype diversity, quantified by
the Shannon entropy (in natural log units) of the true haplotype frequency
distribution, on the performance of the EM algorithm.  We simulated pooled
100bp paired-end sequence data from 162 haplotypes at 200x coverage in a 200kb
region.  We generated the haplotype frequencies using symmetric Dirichlet
distributions with parameter values ranging from .005 to 10, for a total of 550
replicates, which were binned by Shannon entropy (Figure \ref{entropy}).  We
found that the EM algorithm performs best for low entropy frequency
distributions, where there are a few haplotypes at high frequency.  Performance
degrades as the entropy increases, with a slight improvement for high-entropy
(nearly uniform) distributions.  This behavior can be explained by the fact
that missing information leads to uniform estimates, which will give better
results for near-uniform distributions.

\subsection*{Effects of Inaccurate Base Quality Score Reporting}

\begin{figure}[h]
\begin{center}
\includegraphics[scale=0.33]{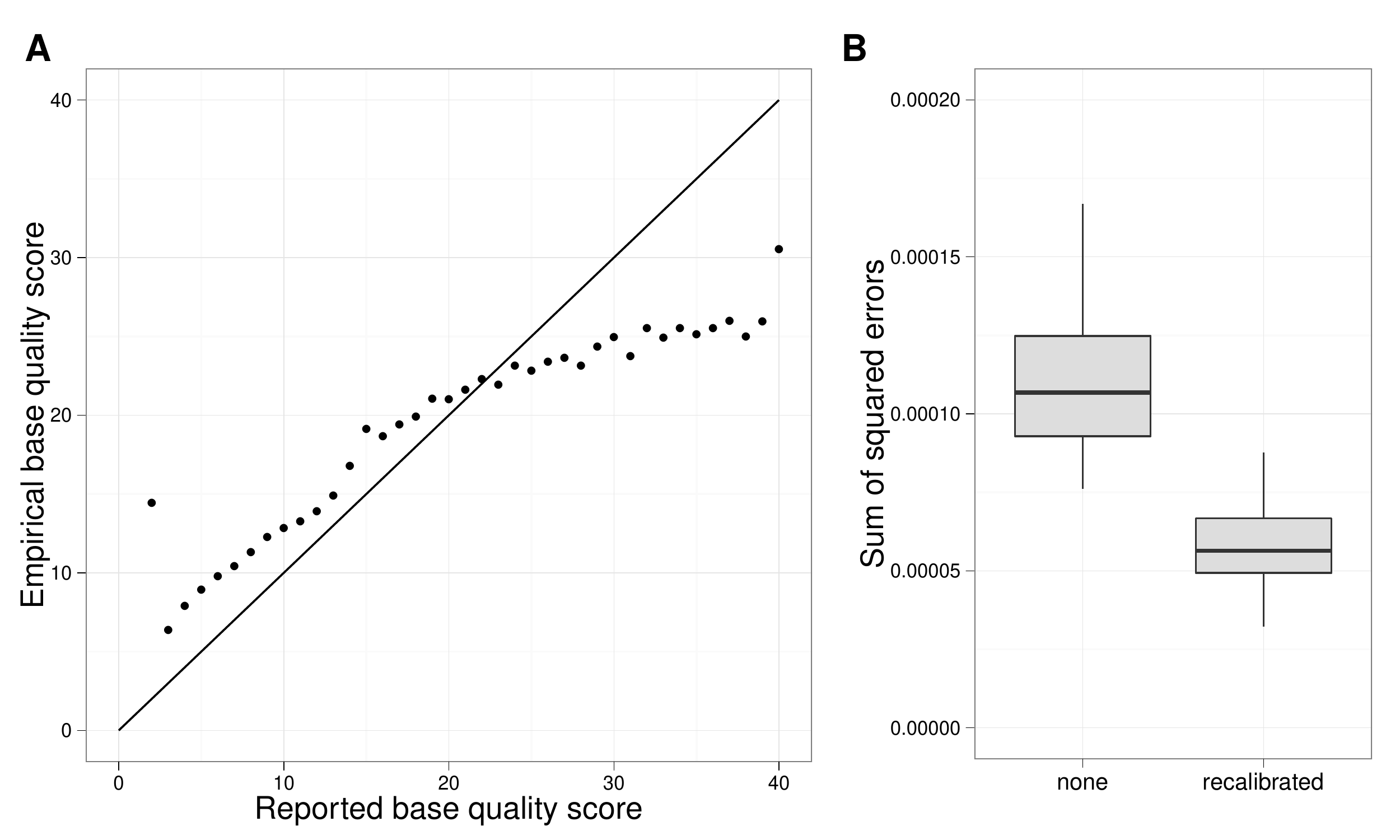}
\end{center}
\caption{A) Reported base quality scores do not match empirical scores
    calculated from real data using monomorphic sites.  B) Recalibration of
    base quality scores using monomorphic sites improves performance.  The
    EM algorithm was run with and without base
    quality score recalibration on simulated pooled 100bp paired-end sequence data from
    162 haplotypes at 200x coverage in a 200kb region.  Sequence errors in the 
    simulated data were introduced with probabilities given by the empirical error rates.}
\label{bqi}
\end{figure}

The computation of haplotype likelihoods is dependent on the correct reporting
and interpretation of base quality scores.  By looking at monomorphic sites in
our experimental data sets, we calculated an observed error rate $P_{obs}(\text{error})$ 
for each possible base quality score, which maps to an empirical base quality score $q_{obs}$
according to: 
\begin{align*}
    q_{obs} = -10 \: \log_{10} P_{obs}(\text{error})
\end{align*}
We observed that the reported base quality scores in our experimental data sets
were consistently inaccurate (Figure \ref{bqi}A).  This motivated the
development of a recalibration method to correct for inaccurate reporting of
base quality scores (see Methods).

In order to test our recalibration method, we simulated data sets (162
haplotypes, 100bp paired-end reads, 200kb region, 200x coverage) using the
empirical error rate for each base quality score to generate sequence read
errors.  For each of 100 replicates, we ran the EM algorithm with and without
recalibration of the base quality scores.  We found the algorithm has higher 
accuracy with the base quality score recalibration (Figure \ref{bqi}B).

\subsection*{Random Initial Estimates To Avoid Local Maxima}

\begin{figure}[h]
\begin{center}
\includegraphics[scale=0.45]{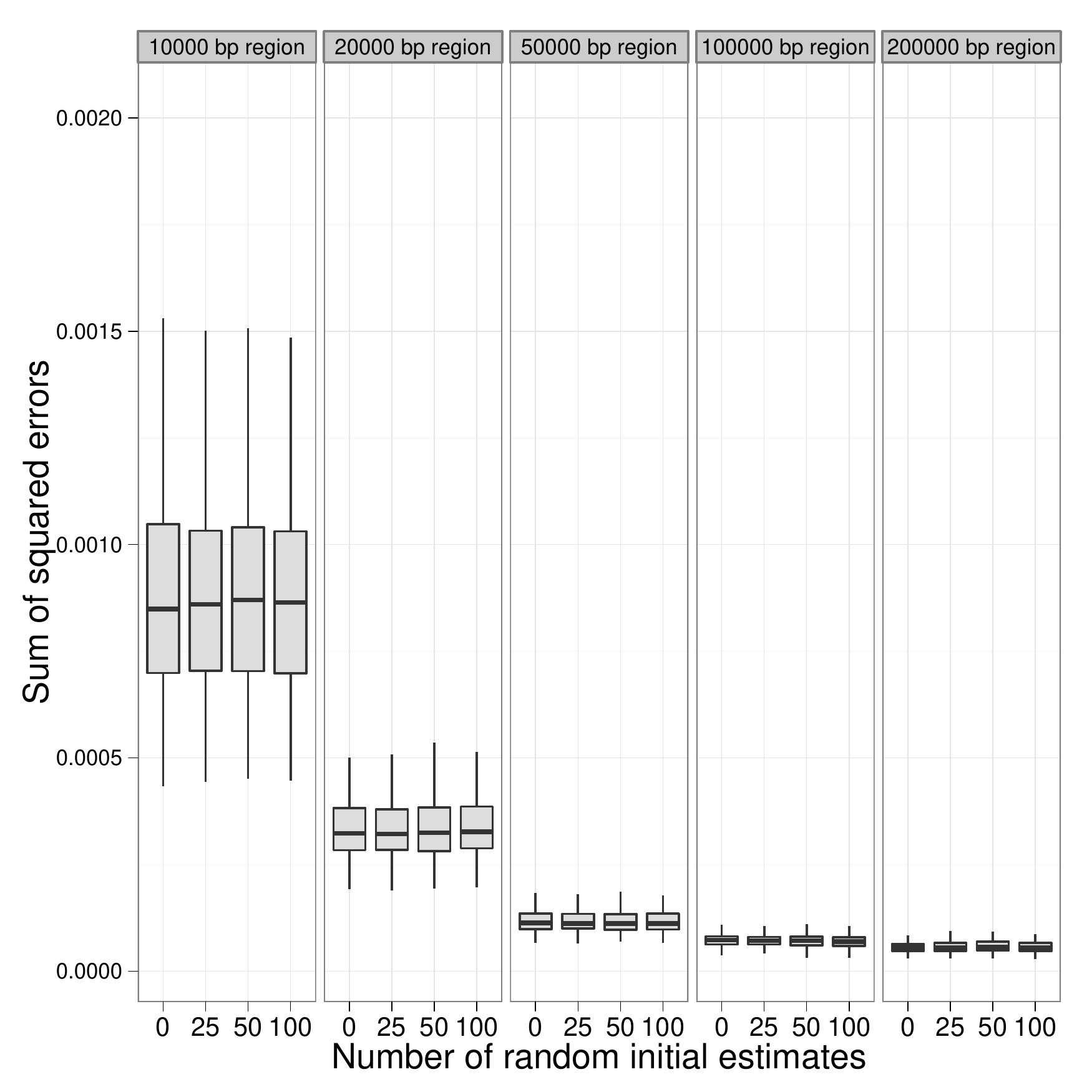}
\end{center}
\caption{The EM algorithm shows no improvement from using multiple random
    initial estimates, showing that it converges to the global maximum reliably.
    The EM algorithm was run on simulated pooled 100bp paired-end sequence data from
    162 haplotypes at 200x coverage, over a range of region widths, with varying 
    numbers of random initial estimates.}
\label{random_starts}
\end{figure}

We investigated the possibility that the EM algorithm could converge to
non-global local maxima on the likelihood surface.  We simulated data sets (162
haplotypes, 100bp paired-end reads, 200x coverage, empirical error rates) over
a range of region sizes from 10kb to 200kb, starting from a uniform initial
estimate in addition to a varying number of random initial estimates (0, 25,
50, 100), with 100 replicates for each combination.  We found that running the
algorithm multiple times with random initial estimates did not improve
performance, indicating that the EM algorithm finds the global maximum reliably
starting from a uniform initial estimate.

\section*{Discussion} % MBE: can be combined with Results

We have presented a new method for estimating the frequencies of known
haplotypes from pooled sequence data, using the haplotype information contained
in individual sequence reads.

We showed that the method outperforms methods based on allele frequencies, as
well as simple methods using sequence data.  Using data from larger genomic
regions improves the accuracy of the estimate.  Increased coverage and longer
read lengths also improve the performance of the algorithm.  The method
generally performs better for haplotype frequency distributions with lower
entropy than those with higher entropy.  The method incorporates uncertainty in
the sequence reads by using the reported base quality scores.  Recalibration of
base quality scores using monomorphic sites in the pooled data leads to better
performance.

The method relies on the probabilistic assignment of sequence reads to the
known haplotypes.  The method works best when the SNP density (the number of SNPs per
base pair) is high enough so that individual read pairs will cover multiple
SNPs.  In the DGRP Drosophila strains, the SNP density is $\sim 1/20$ SNP/bp so
that 100bp paired-end reads contain on average 10 SNPs per pair, which is
sufficient for this probabilistic assignment.  As sequence read lengths
increase with advances in sequencing technology, we anticipate that this method
will be useful for a wide variety of organisms.

This method has immediate application in the analysis of pooled data from
artificial selection experiments where the founding haplotypes are known.  In
essence, by using information from the founding population, we are able to
infer haplotype information about the final pooled population, which has
previously only been available when individuals have been sequenced separately.
This haplotype information can then be used for various purposes (e.g. to look
for signatures of selection).  

It should be noted that in the experimental evolution setting, the haplotype
frequency estimates obtained are local, and will vary across the genome
due to recombination over the course of the experiment.  In the case where a
recombination has occurred within the region under consideration, nearly all
sequence reads coming from the recombined haplotype will come from one or the
other of the two original haplotypes.  Because of this, reads from the
recombined haplotype will contribute to the frequency estimates of both of the
original haplotypes, with the exact proportion determined by the location of
the recombination point within the region.  The few reads that span a
recombination point will have very low haplotype likelihoods, and will
contribute neglibly to the final estimate.  In practice, one can choose the
width of the genomic region used for the estimation to be smaller than the
expected length scale of recombination.  For example, in Drosophila selection
experiments lasting 25 generations, we expect to see recombination breakpoints
at a scale of $\approx 1 \text{mb}$, whereas our method obtains very accurate results
with much smaller regions of $\approx 100 \text{ kb}$.

In addition to applications in experimental evolution, we anticipate that our
method will find application in the estimation of proportions of known pathogen
strains in naturally pooled samples (e.g. blood samples), as well as in
metagenomics contexts.  Haplotype frequency estimation from pools may also be useful in QTL mapping studies. In
 studies with recombinant inbred lines, several generations of inbreeding are carried out with lines that are derived from mixed populations founded by multiple strains, and the parent of origin for each segment in each inbred line is inferred and correlated with phenotypic trait values.  As an alternative, it may be possible to perform haplotype frequency estimation from pooled
sequencing of the mixed populations directly and map traits by correlating haplotype frequencies with trait values.  

The software implementing the method, \texttt{harp}, is open source and
available for download, and can be easily integrated into existing analysis
pipelines.

\section*{Acknowledgments}
This work was supported by the National Institutes of Health 
(Training Grant in Genomic Analysis and Interpretation T32 HG002536 for D.K.,  
R01 GM053275 for J.N., and R01 GM098614 for T.T.) and by the NSF (EF-0928690 for J.N.) and the Searle Scholars Program (J.N.).  
We thank Ken Lange for his suggestions on the standard error calculations.

%
% bibliography
%

\bibliographystyle{mbe}
\bibliography{harp_paper}

\begin{thebibliography}{35}
\providecommand{\natexlab}[1]{#1}

\bibitem[{Burke et~al.(2010)Burke, Dunham, Shahrestani, Thornton, Rose, and
  Long}]{Burke2010}
Burke MK, Dunham JP, Shahrestani P, Thornton KR, Rose MR, Long AD. 2010.
\newblock {Genome-wide analysis of a long-term evolution experiment with
  Drosophila.}
\newblock Nature. 467:587--90.

\bibitem[{Cheesman et~al.(2003)Cheesman, de~Roode, Read, and
  Carter}]{Cheesman2003}
Cheesman SJ, de~Roode JC, Read AF, Carter R. 2003.
\newblock {Real-time quantitative PCR for analysis of genetically mixed
  infections of malaria parasites: technique validation and applications}.
\newblock Molecular and Biochemical Parasitology. 131:83--91.

\bibitem[{Cutler and Jensen(2010)}]{Cutler2010}
Cutler DJ, Jensen JD. 2010.
\newblock {To pool, or not to pool?}
\newblock Genetics. 186:41--3.

\bibitem[{Dempster, Laird and Rubin(1977)Dempster, Laird, and
  Rubin}]{Dempster1977}
Dempster A, Laird N, Rubin D. 1977.
\newblock {Maximum Likelihood from Incomplete Data via the EM Algorithm}.
\newblock Journal of the Royal Statistical Society: Series B (Methodological).
  39:1--38.

\bibitem[{Earley and Jones(2011)}]{Earley2011}
Earley EJ, Jones CD. 2011.
\newblock {Next-generation mapping of complex traits with phenotype-based
  selection and introgression.}
\newblock Genetics. 189:1203--9.

\bibitem[{Futschik and Schl\"{o}tterer(2010)}]{Futschik2010}
Futschik A, Schl\"{o}tterer C. 2010.
\newblock {The next generation of molecular markers from massively parallel
  sequencing of pooled DNA samples.}
\newblock Genetics. 186:207--18.

\bibitem[{Gasbarra et~al.(2009)Gasbarra, Kulathinal, Pirinen, and
  Sillanp\"{a}\"{a}}]{Gasbarra2009}
Gasbarra D, Kulathinal S, Pirinen M, Sillanp\"{a}\"{a} MJ. 2009.
\newblock {Estimating haplotype frequencies by combining data from large DNA
  pools with database information.}
\newblock IEEE/ACM transactions on computational biology and bioinformatics /
  IEEE, ACM. 8:36--44.

\bibitem[{Hastings, Nsanzabana and Smith(2010)Hastings, Nsanzabana, and
  Smith}]{Hastings2010}
Hastings IM, Nsanzabana C, Smith Ta. 2010.
\newblock {A comparison of methods to detect and quantify the markers of
  antimalarial drug resistance.}
\newblock The American journal of tropical medicine and hygiene. 83:489--95.

\bibitem[{Hastings and Smith(2008)}]{Hastings2008}
Hastings IM, Smith Ta. 2008.
\newblock {MalHaploFreq: a computer programme for estimating malaria haplotype
  frequencies from blood samples.}
\newblock Malaria journal. 7:130.

\bibitem[{Huang et~al.(2012)Huang, Richards, Carbone et~al.}]{Huang2012}
Huang W, Richards S, Carbone Ma, et~al. (25 co-authors). 2012.
\newblock {Epistasis dominates the genetic architecture of Drosophila
  quantitative traits}.
\newblock Proceedings of the National Academy of Sciences. pp. 1--7.

\bibitem[{Hunt et~al.(2005)Hunt, Fawcett, Carter, and Walliker}]{Hunt2005}
Hunt P, Fawcett R, Carter R, Walliker D. 2005.
\newblock {Estimating SNP proportions in populations of malaria parasites by
  sequencing: validation and applications.}
\newblock Molecular and biochemical parasitology. 143:173--82.

\bibitem[{Ito et~al.(2003)Ito, Chiku, Inoue, Tomita, Morisaki, Morisaki, and
  Kamatani}]{Ito2003}
Ito T, Chiku S, Inoue E, Tomita M, Morisaki T, Morisaki H, Kamatani N. 2003.
\newblock {Estimation of Haplotype Frequencies, Linkage-Disequilibrium
  Measures, and Combination of Haplotype Copies in Each Pool by Use of Pooled
  DNA Data}.
\newblock American journal of human genetics. pp. 384--398.

\bibitem[{Kirkpatrick et~al.(2007)Kirkpatrick, Armendariz, Karp, and
  Halperin}]{Kirkpatrick2007}
Kirkpatrick B, Armendariz CS, Karp RM, Halperin E. 2007.
\newblock {HAPLOPOOL: improving haplotype frequency estimation through DNA
  pools and phylogenetic modeling.}
\newblock Bioinformatics (Oxford, England). 23:3048--55.

\bibitem[{Kofler et~al.(2011)Kofler, Orozco-terWengel, {De Maio}, Pandey,
  Nolte, Futschik, Kosiol, and Schl\"{o}tterer}]{Kofler2011}
Kofler R, Orozco-terWengel P, {De Maio} N, Pandey RV, Nolte V, Futschik A,
  Kosiol C, Schl\"{o}tterer C. 2011.
\newblock {PoPoolation: a toolbox for population genetic analysis of next
  generation sequencing data from pooled individuals.}
\newblock PloS one. 6:e15925.

\bibitem[{Kuk, Zhang and Yang(2009)Kuk, Zhang, and Yang}]{Kuk2009}
Kuk AYC, Zhang H, Yang Y. 2009.
\newblock {Computationally feasible estimation of haplotype frequencies from
  pooled DNA with and without Hardy-Weinberg equilibrium.}
\newblock Bioinformatics. 25:379--86.

\bibitem[{Lange(2010)}]{Lange2010}
Lange K. 2010.
\newblock Numerical Analysis for Statisticians (Statistics and Computing).
\newblock Springer, 2nd ed. edition.

\bibitem[{Ley et~al.(2006)Ley, Turnbaugh, Klein, and Gordon}]{Ley2006}
Ley R, Turnbaugh PJ, Klein S, Gordon JI. 2006.
\newblock {Human gut microbes associated with obesity}.
\newblock Nature. 444.

\bibitem[{Li et~al.(2009)Li, Handsaker, Wysoker, Fennell, Ruan, Homer, Marth,
  Abecasis, and Durbin}]{Li2009}
Li H, Handsaker B, Wysoker A, Fennell T, Ruan J, Homer N, Marth G, Abecasis G,
  Durbin R. 2009.
\newblock {The Sequence Alignment/Map format and SAMtools.}
\newblock Bioinformatics (Oxford, England). 25:2078--9.

\bibitem[{Li et~al.(2007)Li, Foulkes, Yucel, and Rich}]{Li2007}
Li X, Foulkes AS, Yucel RM, Rich SM. 2007.
\newblock {An expectation maximization approach to estimate malaria haplotype
  frequencies in multiply infected children.}
\newblock Statistical applications in genetics and molecular biology.
  6:Article33.

\bibitem[{Long et~al.(2011)Long, Jeffares, Zhang, Ye, Nizhynska, Ning,
  Tyler-Smith, and Nordborg}]{Long2011}
Long Q, Jeffares DC, Zhang Q, Ye K, Nizhynska V, Ning Z, Tyler-Smith C,
  Nordborg M. 2011.
\newblock {PoolHap: inferring haplotype frequencies from pooled samples by next
  generation sequencing.}
\newblock PloS one. 6:e15292.

\bibitem[{Mackay et~al.(2012)Mackay, Richards, Stone et~al.}]{Mackay2012}
Mackay TFC, Richards S, Stone Ea, et~al. (52 co-authors). 2012.
\newblock {The Drosophila melanogaster Genetic Reference Panel}.
\newblock Nature. 482:173--178.

\bibitem[{Niu(2004)}]{Niu2004}
Niu T. 2004.
\newblock {Algorithms for inferring haplotypes.}
\newblock Genetic epidemiology. 27:334--47.

\bibitem[{Nuzhdin et~al.(2007)Nuzhdin, Harshman, Zhou, and
  Harmon}]{Nuzhdin2007}
Nuzhdin SV, Harshman LG, Zhou M, Harmon K. 2007.
\newblock {Genome-enabled hitchhiking mapping identifies QTLs for stress
  resistance in natural Drosophila.}
\newblock Heredity. 99:313--21.

\bibitem[{Orozco-terWengel et~al.(2012)Orozco-terWengel, Kapun, Nolte, Kofler,
  Flatt, and Schl\"{o}tterer}]{Orozco-terWengel2012}
Orozco-terWengel P, Kapun M, Nolte V, Kofler R, Flatt T, Schl\"{o}tterer C.
  2012.
\newblock {Adaptation of Drosophila to a novel laboratory environment reveals
  temporally heterogeneous trajectories of selected alleles.}
\newblock Molecular ecology. .

\bibitem[{Pe'er and Beckmann(2003)}]{Pe'er2003}
Pe'er I, Beckmann JS. 2003.
\newblock {Resolution of haplotypes and haplotype frequencies from SNP
  genotypes of pooled samples}.
\newblock Proceedings of the seventh annual international conference on
  Computational molecular biology - RECOMB '03. pp. 237--246.

\bibitem[{Pirinen(2009)}]{Pirinen2009}
Pirinen M. 2009.
\newblock {Estimating population haplotype frequencies from pooled SNP data
  using incomplete database information.}
\newblock Bioinformatics. 25:3296--302.

\bibitem[{Sabeti et~al.(2007)Sabeti, Varilly, Fry et~al.}]{Sabeti2007}
Sabeti PC, Varilly P, Fry B, et~al. (248 co-authors). 2007.
\newblock {Genome-wide detection and characterization of positive selection in
  human populations.}
\newblock Nature. 449:913--8.

\bibitem[{Takala et~al.(2006)Takala, Smith, Stine, Coulibaly, Thera, Doumbo,
  and Plowe}]{Takala2006}
Takala SL, Smith DL, Stine OC, Coulibaly D, Thera Ma, Doumbo OK, Plowe CV.
  2006.
\newblock {A high-throughput method for quantifying alleles and haplotypes of
  the malaria vaccine candidate Plasmodium falciparum merozoite surface
  protein-1 19 kDa.}
\newblock Malaria journal. 5:31.

\bibitem[{Turner and Miller(2012)}]{Turner2012}
Turner TL, Miller PM. 2012.
\newblock {Investigating natural variation in Drosophila courtship song by the
  evolve and resequence approach.}
\newblock Genetics. 191:633--42.

\bibitem[{Turner et~al.(2011)Turner, Stewart, Fields, Rice, and
  Tarone}]{Turner2011}
Turner TL, Stewart AD, Fields AT, Rice WR, Tarone AM. 2011.
\newblock {Population-based resequencing of experimentally evolved populations
  reveals the genetic basis of body size variation in Drosophila melanogaster.}
\newblock PLoS genetics. 7:e1001336.

\bibitem[{Voight et~al.(2006)Voight, Kudaravalli, Wen, and
  Pritchard}]{Voight2006}
Voight BF, Kudaravalli S, Wen X, Pritchard JK. 2006.
\newblock {A map of recent positive selection in the human genome.}
\newblock PLoS biology. 4:e72.

\bibitem[{Wang, Kidd and Zhao(2003)Wang, Kidd, and Zhao}]{Wang2003}
Wang S, Kidd KK, Zhao H. 2003.
\newblock {On the Use of DNA Pooling to Estimate Haplotype Frequencies}.
\newblock Genetic Epidemiology. 24:74--82.

\bibitem[{Yang et~al.(2003)Yang, Zhang, Hoh, Matsuda, Xu, Lathrop, and
  Ott}]{Yang2003}
Yang Y, Zhang J, Hoh J, Matsuda F, Xu P, Lathrop M, Ott J. 2003.
\newblock {Efficiency of single-nucleotide polymorphism haplotype estimation
  from pooled DNA.}
\newblock Proceedings of the National Academy of Sciences of the United States
  of America. 100:7225--30.

\bibitem[{Zhang, Yang and Yang(2008)Zhang, Yang, and Yang}]{Zhang2008}
Zhang H, Yang HC, Yang Y. 2008.
\newblock {PoooL: an efficient method for estimating haplotype frequencies from
  large DNA pools.}
\newblock Bioinformatics. 24:1942--8.

\bibitem[{Zhou et~al.(2011)Zhou, Udpa, Gersten et~al.}]{Zhou2011}
Zhou D, Udpa N, Gersten M, et~al. (11 co-authors). 2011.
\newblock {Experimental selection of hypoxia-tolerant Drosophila melanogaster.}
\newblock Proceedings of the National Academy of Sciences of the United States
  of America. 108:2349--54.

\end{thebibliography}

%\thebibliography{0}
%
%\bibitem{AR}
%Archibald JM, Roger AJ. 2002.
%Gene conversion and the evolution of euryarchaeal chaperonins:
%a maximum likelihood-based method for detecting conflicting
%phylogenetic signal.
%J. Mol. Evol. 55:232--245.
%
%
%\endthebibliography

\section*{Appendix: Formal EM Calculation}
\bigskip

\paragraph*{Expectation step:} We calculate the expectation of the
complete data log-likelihood, where the expectation is taken over the posterior
distribution of the missing data given the observed data and the current
estimate.  Recall that $p_{j,h} \:=\: P(\eta_j = h|r_j, f^{(i)})$ is the probability
that read $r_j$ came from haplotype $h$, given our current haplotype frequency estimate
$f^{(i)}$.

\begin{align*}
    Q(f|f^{(i)}) &= E_{\eta|r,f^{(i)}} [ \log L(f|\eta,r) ] \\
                 &= E_{\eta|r,f^{(i)}} \sum_j \log P(\eta_j, r_j | f) \\
                 &= \sum_j E_{\eta_j|r_j,f^{(i)}} \log P(\eta_j, r_j | f) \\
                 &= \sum_j \sum_h P(\eta_j=h | r_j, f^{(i)}) \log P(h, r_j | f) \\
                 &= \sum_j \sum_h p_{j,h} [ \log P(r_j|h) + \log P(h|f) ] \\
                 &= \sum_j \sum_h p_{j,h} \log f_h + C \\
                 &= \log \prod_h f_h^{\sum_j p_{j,h}} + C
\end{align*}
where $C$ is a constant independent of $f$.

\paragraph*{Maximization step:} Our next estimate $f^{(i+1)}$ is given by the
$f$ that maximizes the expected log-likelihood:

\begin{equation*}
    f^{(i+1)} = \argmax_f Q(f \, | \, f^{(i)})
\end{equation*}

First note that the function $R(f) = \prod_h f_h^{\alpha_h}$ is maximized by $f_h =
\alpha_h / \sum_i \alpha_i$.  (For example, the maximum likelihood estimator
for the parameters of a multinomial distribution is given by the vector of count
proportions.)

Since $\log$ is monotonic and 
\begin{equation*}
    \sum_h \sum_j p_{j,h} \,=\, \sum_j 1 \,=\, N \text{ (the number of reads)},
\end{equation*}
$Q(f|f^{(i)})$ is maximized when, for all $h$:
\begin{equation*}
f_h = \frac{\sum_j p_{j,h}}{N}.
\end{equation*}

In other words, our next estimate $f^{(i+1)}$ is given by the average of the posterior vectors:
\begin{equation*}
    f^{(i+1)} = \frac{\sum_j p_j}{N}
\end{equation*}

\bigskip

\section*{Appendix: Calculation of Standard Errors}

We use general properties of the EM algorithm and maximum likelihood estimators
to calculate standard errors for our haplotype frequency estimates, following
\cite{Lange2010}.  For brevity, we let $L(f)$ be the log-likelihood of $f$.  Our
strategy to calculate standard errors of our maximum likelihood estimate
$\hat{f}$ is as follows:
\begin{enumerate}
    \item estimate the observed information $I = -d^2 L(\hat{f})$
    \item $\hat{f}$ is asymptotically normal with covariance matrix $I^{-1}$, so
        the standard error estimates are the square roots of the diagonals of $I^{-1}$
\end{enumerate}
One slight complication is that our estimate $\hat{f}$ is subject to a linear constraint
(the frequencies must sum to 1).  Below, we also show how to adjust this calculation to handle
this constraint.

\subsection*{Calculating $d^2 L$}

Let $g$ be the minorizing function for the EM algorithm:
\begin{align*}
    g(f | f_0) &= Q(f | f_0) + L(f_0) - Q(f_0 | f_0) \\
\end{align*}
Then $g$ satisfies the relations for all $f, f_0$:
\begin{align*}
    g(f | f_0) &\leq L(f) \\
    g(f_0 | f_0) &= L(f_0) \\
\end{align*}

Note that $\nabla g(f | f_0) = \nabla Q(f | f_0)$.
Also, $L(f) - g(f | f_0)$ is minimized $f = f_0$, so $\nabla L(f) - \nabla g(f|f_0) = 0$ at $f = f_0$. 
This means that $\nabla L(f_0) = \nabla Q(f_0 | f_0)$.  Note
that $\nabla Q(f_0 | f_0)$ is the gradient $\nabla Q(f|f_0)$ computed as a function of $f$, then evaluated
at $f = f_0$.

In summary, we can write the score function $S = (S_1, \ldots, S_H)'$, defined to be
the gradient of the log-likelihood, as:
\begin{align*}
    S(f) = \nabla L(f) = \nabla Q(f|f)
\end{align*}

Since $dS = d^2 L$, we need to find the partial deriviatives of $S$.

Recall that we have:
\begin{align*}
    Q(f|f_0) &= \sum_h (\sum_j p_{j,h}) \log f_h
\end{align*}
where $p_{j,h}$ is the haplotype posterior vector, representing the probability that
read $r_j$ came from haplotype $h$.

Note that the $p_{j,h}$ depend on $f_0$, but not $f$, so the partial derivatives of $Q(f|f_0)$ have a simple form:
\begin{align*}
    \pd{Q(f|f_0)}{f_h} &= \frac{\sum_j p_{j,h}(f_0)}{f_h} \\
\end{align*}
Now we evaluate at $f = f_0$, and drop the subscript:
\begin{align*}
    S_h(f) =  \frac{\sum_j p_{j,h}(f)}{f_h} \\
\end{align*}
We can now compute partial derivatives of the score function:
\begin{align*}
    \pd{S_h}{f_k} = \frac{1}{f_h} \sum_j \pd{p_{j,h}}{f_k} - \frac{\one{k=h}}{f_h^2} \sum_j p_{j,h} \\
\end{align*}

To compute the partial derivatives of $p_{j,h}$, first write $p_{j,h}$ as:
\begin{align*}
    p_{j,h} = \frac {l_{j,h} f_h}{P_j}
\end{align*}
where $P_j = \sum_h l_{j,h} f_h$ is the total probability of read $r_j$.  
Since $\pd{P_j}{f_k} = l_{j,k}$, we have:
\begin{align*}
    \pd{p_{j,h}}{f_k} &= \frac {l_{j,h}}{P_j} \one{k=h} - l_{j,h} f_h \frac{1}{P_j^2} \pd{P_j}{f_k} \\
                      &= \frac {l_{j,h}}{P_j} \one{k=h} - \frac{l_{j,h} l_{j,k} f_h}{P_j^2} \\ 
\end{align*}

\subsection*{Adjusting for the linear constraint}
We continue to follow \cite{Lange2010} to handle the linear constraint $\sum_h f_h = 1$.  
Let $V = \mathbf{1}_H^t = (1 1 \cdots 1)$ be the row vector with $H$ 1's, so
we can write our constraint as $Vf = 1$.

We let $W$ be a matrix with $H - 1$ column vectors orthogonal to $V$:
\begin{align*}
    W =
    \begin{pmatrix}
        1 & 1 & \cdots & 1 \\
        -1 & 0 & \cdots & 0 \\
        0 & -1 & \cdots & 0 \\
        \vdots & \vdots & \vdots & \vdots \\
        0 & 0 & \cdots & -1 \\
    \end{pmatrix}
\end{align*}

We reparametrize by $\beta$, using the relation $f = \alpha + W \beta$, 
where $\alpha$ satisfies the constraint $V\alpha = 1$.  As a function of $\beta$,
the log-likelihood $L(\alpha + W \beta)$ has observed information $-W^t d^2 L(\alpha + W \beta) W$,
which gives an estimate $\text{Var}(\hat{\beta}) = -[W^t d^2 L(\hat{f}) W]^{-1}$.
This gives the estimate 
$\text{Var}(\hat{f}) = \text{Var}(W \hat{\beta}) 
    = - W \: [W^t \: d^2 L(\hat{f}) \: W ]^{-1} \: W^t$, 
where $d^2 L(\hat{f})$ was estimated above.

\end{document}